# Title: Facile synthesis of Cu$_2$O nanorods in the presence of NaCl by SILAR method and its characterizations.

Md. Alauddin Hossain[a], Syed Farid Uddin Farhad[b,*],

Nazmul Islam Tanvir[b], Jang Hyo Chang[d], Mohammad Atiqur Rahman[a, c],

Tooru Tanaka[d], Qixin Guo[d], Jamal Uddin[e], Md Abdul Majed Patwary[a, d,*],

[a] *Physical Chemistry Research Laboratory, Department of Chemistry, Comilla University, Cumilla-3506, Bangladesh*
[b] *Energy Conversion and Storage Research Section, Industrial Physics Division, BCSIR Laboratories, Dhaka-1205, Bangladesh*
[c] *Department of Chemistry, Kumamoto University, Kumamoto, 860-8555, Japan*
[d] *Department of Electrical and Electronic Engineering, Saga University, Saga, 840-8502, Japan*
[e] *Center for Nanotechnology, Department of Natural Sciences, Coppin State University, Baltimore, MD, USA*

*Corresponding Author: mamajedp@gmail.com (M. A. M. Patwary),* https://orcid.org/0000-0003-4680-957X); *s.f.u.farhad@bcsir.gov.bd (S. F. U. Farhad)*

**Abstract:**

Cu$_2$O nanorods have been deposited on soda-lime glass (SLG) substrates by the modified SILAR technique by varying the concentration of NaCl electrolyte into the pre-cursor complex solution. The structural, electrical, and optical properties of synthesized Cu$_2$O nanorod films have been studied by a variety of characterization tools. Structural analyses by XRD confirmed the polycrystalline Cu$_2$O phase with (111) preferential growth. Raman scattering spectroscopic measurements conducted at room temperature also showed characteristic peaks of the pure Cu$_2$O phase. The surface resistivity of the Cu$_2$O nanorod films decreased from 15,142 to 685 $\Omega$.cm with the addition of NaCl from 0 to 4 mmol, and then exhibited an opposite trend with further addition of NaCl. The optical bandgap of the synthesized Cu$_2$O nanorod films was observed as 1.88~2.36 eV, while the temperature-dependent activation energies of the Cu$_2$O films were measured as about 0.14~0.21 eV. SEM morphologies demonstrated Cu$_2$O nanorod as well as closely packed spherical grains with the alteration of NaCl concentration. The Cu$_2$O phase of nanorods was found stable up to 230 ℃ corroborating the optical bandgap results of the same. The film fabricated in presence of 4 mmol of NaCl showed the lowest resistivity and activation energy as well comparatively uniform nanorod morphology. Our studies demonstrate that the nominal presence of NaCl electrolytes in the pre-cursor solutions has a significant impact on the physical properties of Cu$_2$O nanorod films which could be beneficial in optoelectronic research.

**Introduction:**

Cuprous oxide (Cu$_2$O) is a *p*-type intrinsic semiconductor due to copper vacancies in the crystal lattice with the bandgap of ~ 2.17 eV [1, 2] having several promising advantages such as high abundance, low-cost production, visible-light harvesting, and non-toxicity. Cu$_2$O has attracted interest as a good candidate material for photocatalysis [3, 4] chemo sensing [5, 6], electrode materials in lithium-ion batteries [7, 8], photovoltaics [9,10] and photoelectrochemical water splitting [4, 11]. Still, there are numerous methodological challenges in the research and application of Cu$_2$O including the consistent synthesis of nanostructured Cu$_2$O materials as well as the expected formation of Cu° interlayers at *p-n* heterojunctions [12, 13]. Moreover, the maximum theoretical limit of the efficiency of single-junction Cu$_2$O is as high as 20 % under air mass (AM) 1 solar illumination [14], which is far from the achieved results. Recently, Minami et. al. fabricated heterojunction solar cells by inserting an *n*-type zinc-germanium-oxide (Zn$_{1-X}$Ge$_X$O) thin-film between an Al-doped ZnO thin film and a *p*-type Na-doped Cu$_2$O (Cu$_2$O: Na) sheet prepared by thermally oxidized Cu sheets and reported the conversion efficiency of the cell as 8.1% [9]. On the other hand, Ci et. al. fabricated Cl-doped *n*-type Cu$_2$O films by chemical bath deposition by using CuSO$_4$ solution with the addition of CuCl$_2$ as a Cl$^-$ source [15]. Therefore, it looks crucial to study the influence of Na and Cl individually or the electrolytic behavior of NaCl in a broad spectrum on Cu$_2$O film deposition.

There are various methods to synthesis $Cu_2O$ thin films such as atomic layer deposition (ALD) [16], electrochemical deposition [17 - 20], metal-organic chemical vapor deposition (MOCVD) [21], molecular beam epitaxy [22], successive ionic layer adsorption and reaction (SILAR) [23 - 25], the direct oxidation of Cu sheets [26], sputtering [27], vapor phase epitaxy [28], and sol–gel technique [29]. But toward the preparation of nanostructured $Cu_2O$, the electrodeposition of $Cu_2O$ thin films allows the greatest control yet [24 - 27]. To the best of our knowledge, there is no study about the SILAR deposition of $Cu_2O$ nanorod films till now. Moreover, chemical bath optimization for epitaxial growth of $Cu_2O$ nanorods from surfaces has not yet been reported [30].

SILAR is basically one of the most simple and cost-effective methods since it does not require sophisticated apparatus as required in electrodeposition or other methods. In this study, a modified SILAR method was used to synthesize $Cu_2O$ nanostructured thin films. In our prior report, we have shown that the modification [23-25] of SILAR by eliminating the rinsing steps during the growth of pure $Cu_2O$ thin films and named the method as the modified (m) SILAR technique. Therefore, we synthesized $Cu_2O$ nanorods on SLG substrates by m-SILAR method in the nominal presence of NaCl into the pre-cursor solution complex. The objective of this work was to study the impact of the NaCl electrolyte concentrations on the growth of $Cu_2O$ nanorod films through the investigation of their structural, morphological, optical, and electrical properties for photovoltaic applications.

## 2.1 Materials

In this work, sodium thiosulfate pentahydrate ($Na_2S_2O_3.5H_2O$; Scharlau: purity ~99.0%), copper (II) sulfate pentahydrate ($CuSO_4.5H_2O$; Merck Millipore: purity ~99.0%), sodium hydroxide (NaOH; Active fine chemicals: purity ~98.0%) and sodium chloride (NaCl; Merck Millipore: purity ~99%) were collected from the local market and used without further refinement. Soda-lime glass (SLG) microscopy slides (25×25×1 $mm^3$) were used as substrates to deposit copper oxide thin films.

## 2.2 Synthesis of Copper (I) Oxide Thin Films

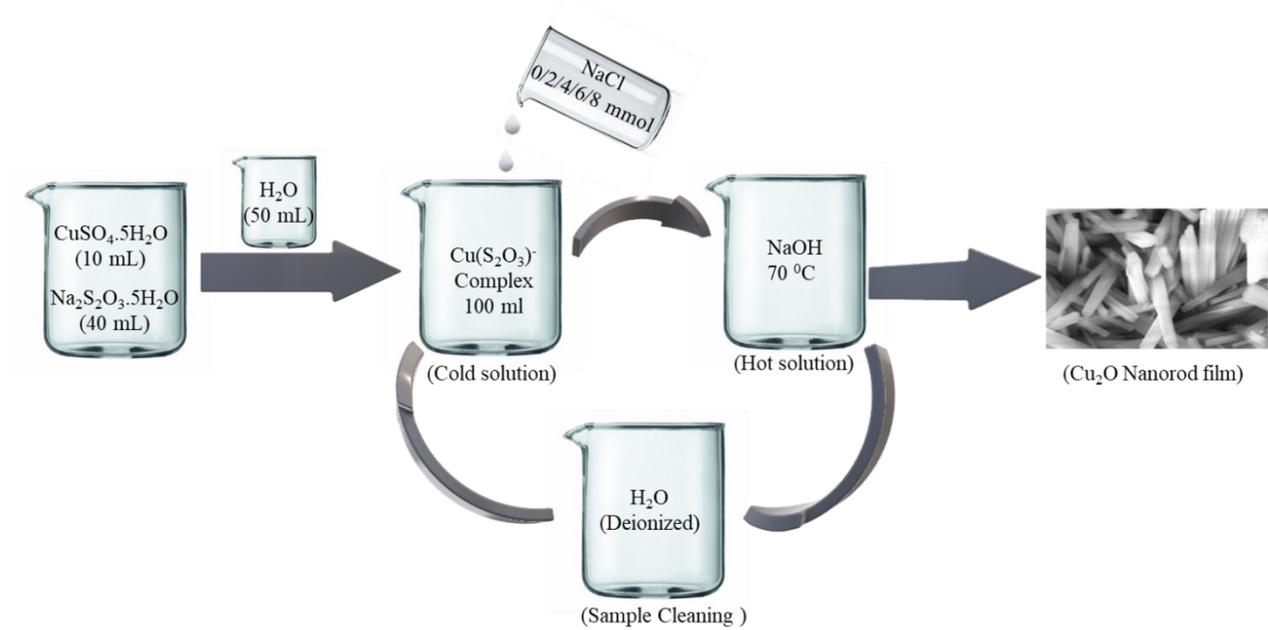

Fig. 1: Synthesis of copper (I) oxide nanorod thin films

$Cu_2O$ thin films were deposited on SLG substrates as shown in Fig. 1 by using the same method described in our previous work [23]. Briefly, the SLG substrates were initially cleaned by detergent to remove loosely attached visible dust particles. Then, the substrates were successively cleaned in an ultrasonic water bath through deionized (DI) water, ethanol, toluene, and isopropanol for 15 min in each case. Prior to the film deposition, 10 mL 1M copper (II) sulfate and 40 mL 1M sodium thiosulfate solution was added into 100 mL volumetric flask until the colorless solution of copper-thiosulfate complex appeared. Then, 2 mmol NaCl electrolyte was added into the same flask and after shaking the remaining portion was filled with DI water. This complex solution was labeled as a cold solution. 2M NaOH solution was kept at 70 °C (hot solution). After that, the SLG substrate was alternatively immersed in hot and cold solution respectively for 2-3 s each and had completed one SILAR cycle. This process was repeated up to 40 immersion cycles. $OH^-$ and $Cu^+$ ions were adsorbed on the substrate respectively when immersed in a hot and cold solution. Consequently, $Cu_2O$ thin films were deposited on the substrate due to the following chemical reactions:

$$Cu(S_2O_3)^- \rightarrow Cu^+ + S_2O_3^{2-}$$

$$2Cu^+ + OH^- \rightarrow Cu_2O + H_2O$$

After deposition, this as-made sample was washed through DI water to eliminate loosely bound particles and dried naturally in the laboratory ambient. Other samples were prepared in a similar way and stored safely in an air-tight sample box for future characterization purposes. The growth mechanism of $Cu_2O$ nanorods are discussed later in the surface morphology section with the support of SEM micrographs.

**2.3 Characterization Process**

The crystal structure and phase present in the samples were examined through X-ray diffraction (XRD) spectrometer (Philips PANalytical X'Pert MRD) under θ - 2θ coupled mode with CuKα radiation source of wavelength, λ = 0.15406 nm as well as sensitive Raman scattering spectrometer (Horiba HR800) where the excitation radiation was 488 nm laser source. Philips XL30 EEG, scanning electron microscope (SEM) was employed to investigate the morphological properties of the samples. Shimadzu UV 2600 ISR Plus UV-VIS-NIR spectrophotometer of wavelength, λ= 220-1400 nm was applied to study the optical response of the deposited samples. A homemade 4-point collinear probe coupled with a Keithley SMU2450 was used to measure the surface resistivity of the samples. Temperature-dependent surface resistivity was measured by air annealing the samples from 30~230 °C through a homemade 2-probe system coupled with a digital multimeter (BK Precision 2704C).

**3 Results and discussion**

**3.1 Crystal structure and phase identification**

The crystal structure and phases present in the samples deposited on SLG substrate have been examined through X-ray diffraction (XRD) spectroscopy under θ − 2θ coupled mode in the range 30° − 70° and the relevant XRD pattern is illustrated in Fig. 2(a). All the samples exhibited distinguished peaks at 2θ ≈ 36.5°, 42.5° and 61.5° respectively which were assigned to (111), (200) and (220) plane of pure cubic phase of $Cu_2O$ only that matches to the High-Quality Inorganic Crystal Structure Database (ICSD) of phase pure $Cu_2O$ (ICSD PDF#180846) [23], and none of the peaks from Cu or CuO phases were present. Fig. 2(a) revealed that all the films were polycrystalline in nature with (111) preferential growth. The intensity of the

(111) plane of Cu$_2$O is increased with increasing the concentration of NaCl electrolyte (2~8 mmol) which indicates the improvement of the crystalline quality of the deposited films [24].

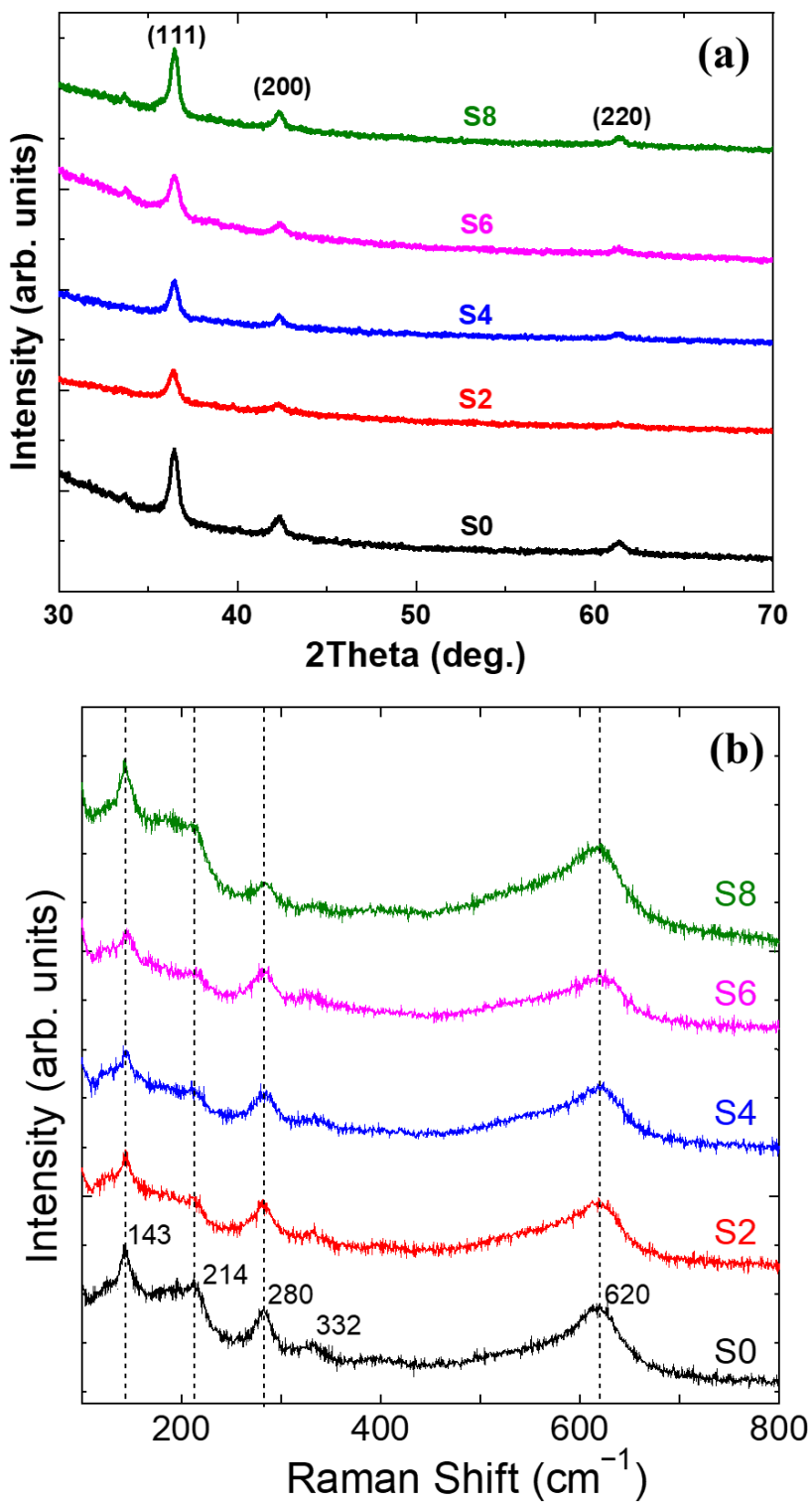

Fig. 2: (a) XRD pattern and (b) Raman spectra of the samples deposited on SLG substrate in presence of NaCl electrolyte with various concentrations

The texture coefficient (TC) was assessed to describe the crystallographic nature of the deposited films by using the ratio, TC (hkl): $\frac{I(111)}{I(111)+I(200)}$, where I(111) and I(200) were the intensity of (111) and (200) planes respectively [23] and the calculated values are shown in Table 1. The average crystallite size (D) was determined by using the Scherrer Formula [31] and it was 14.91~16.51 nm. The values of some other structural parameters are listed in Table 1. The increasing value of TC; similarly shifting of 2θ values to the higher diffraction angle with respect to the reference 2θ value (marked by an asterisk in Table-1 and the respective plot is inserted in Fig. 3(b)) signified the improvement of the crystalline quality of the deposited films in presence of a higher concentration of NaCl electrolyte [32]. Variation of crystallite size and texture coefficient are shown in Fig. 3(a). Although the little amount of NaCl electrolyte (2 mmol) deteriorate the crystalline quality of the film in the case of sample S2 with respect to S0 (zero NaCl), but at higher NaCl concentration, the nano crystallinity of the films was improved which can be seen from Fig. 3(a). In this case, sample S8 showed the highest crystallinity among all the samples having the largest crystallite size (16.51 nm) with the minimum dislocation density and micro strain.

Table 1: Structural parameters of the deposited films

| Sample | Conc. of NaCl (mmol) | 2θ (deg.) | d(111) (nm) | a (nm) | TC(hkl) | FWHM (°) | Crystallite (nm) | Dislocation density $\delta \times 10^{-3}$ (nm$^{-2}$) | Strain $\varepsilon \times 10^{-3}$ |
|---|---|---|---|---|---|---|---|---|---|
| Ref.* | 0 | 36.42 | 0.2465 | 0.4270 | 0.75 | 0.33 | 27.00 | 1.372 | 1.37 |
| S0 | 0 | 36.46 | 0.2462 | 0.4265 | 0.65 | 0.50 | 15.94 | 3.94 | 2.17 |
| S2 | 2 | 36.40 | 0.2466 | 0.4272 | 0.62 | 0.52 | 15.24 | 4.30 | 2.27 |
| S4 | 4 | 36.47 | 0.2462 | 0.4264 | 0.63 | 0.53 | 14.91 | 4.50 | 2.31 |
| S6 | 6 | 36.47 | 0.2462 | 0.4264 | 0.63 | 0.56 | 14.14 | 5.00 | 2.45 |
| S8 | 8 | 36.47 | 0.2462 | 0.4264 | 0.66 | 0.48 | 16.51 | 3.67 | 2.09 |

*$Cu_2O$ powder (purity: 99.99 %)

To further confirm the phases of the deposited films, the sensitive Raman scattering spectroscopy was also used, and the corresponding Raman spectra is given in Fig. 2(b). Raman peaks at 143, 214, 280, 332 and 620 cm$^{-1}$ corresponds to $Cu_2O$ phase only [27, 33, 34] which is perfectly coincided with the demonstrated

XRD results and has alike pattern with the ref [35]. Thus, the addition of NaCl electrolyte to cationic precursor solution unaltered the chemical environment of the deposited films while improving nano crystallinity.

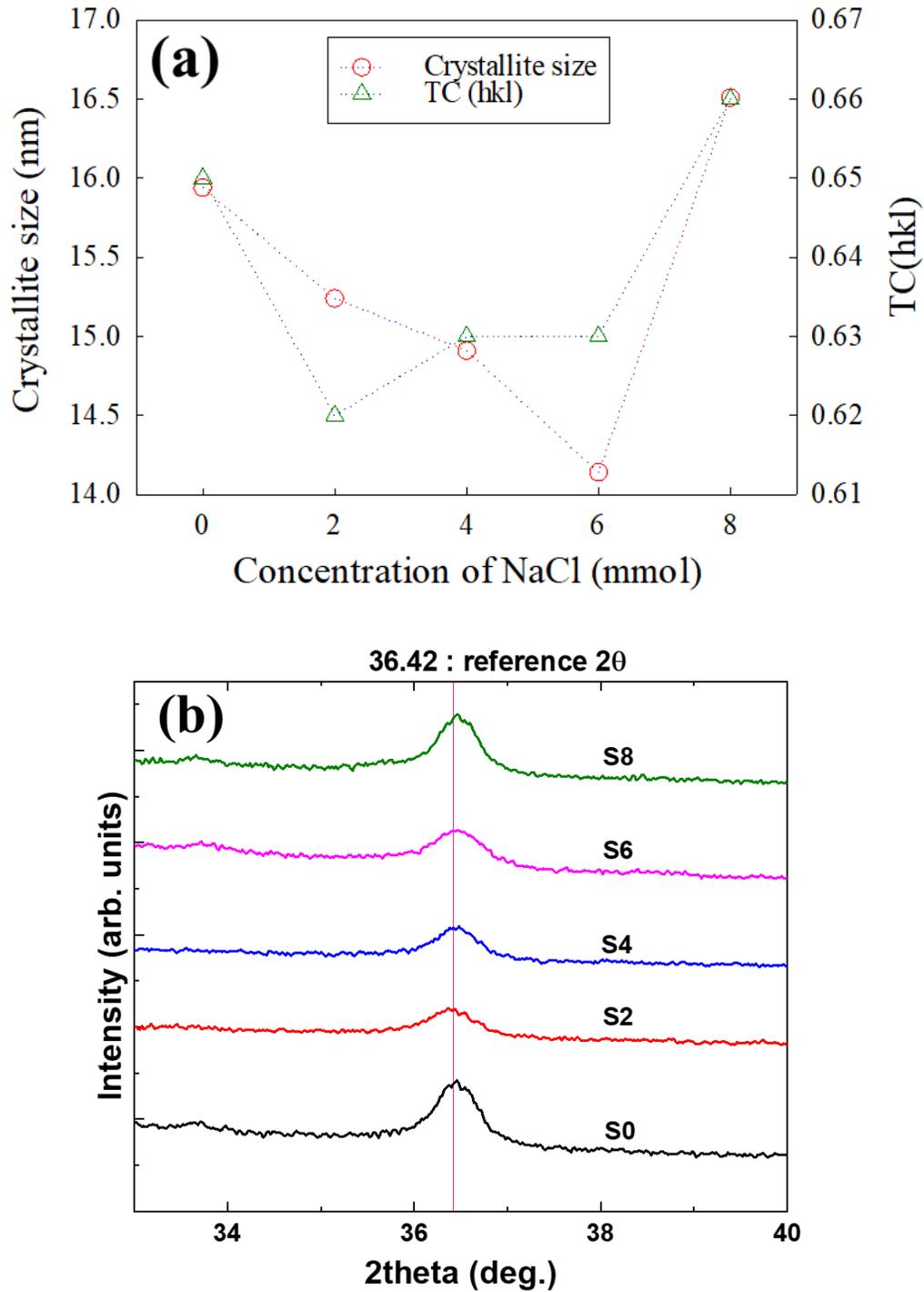

Fig. 3: (a) Variation of crystallite size and texture coefficient with respect to NaCl concentration and (b) shifted 2θ values with respect to the reference one (indicated by the vertical line)

## 3.2 Morphological Analysis:

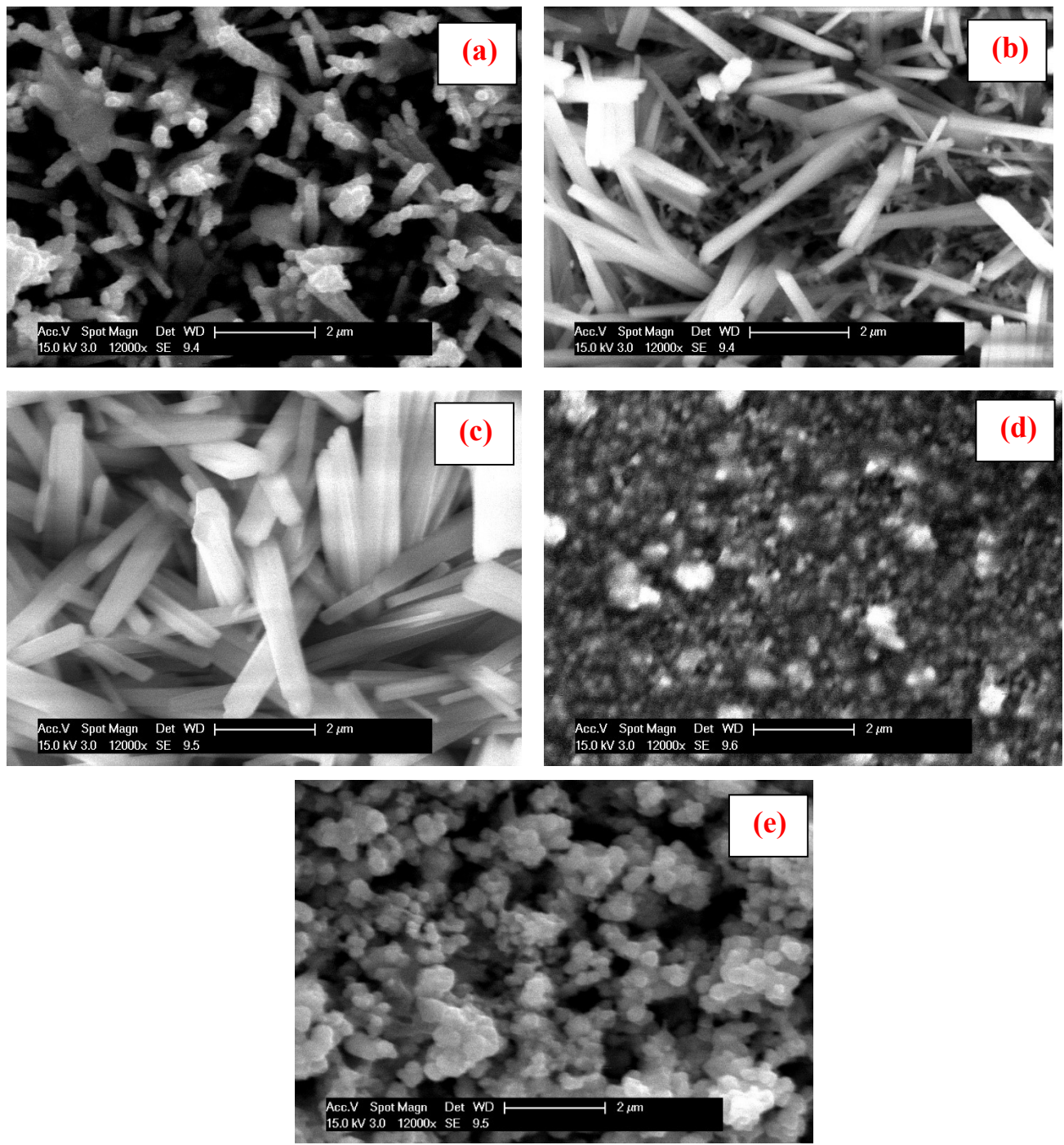

Fig. 4: Surface morphologies of the samples deposited at a) 0 mmol b) 2 mmol c) 4 mmol d) 6 mmol and e) 8 mmol of NaCl electrolyte.

Fig. 4 depicts the SEM micrographs of the samples deposited in presence of NaCl electrolyte at various concentrations. It is observed that the morphologies are crack-free and very well distributed on the substrate surface. The sample deposited without NaCl electrolyte revealed pencil-thin nanorod surface morphology with an overgrown cluster in some regions, as also observed in our previous report [24]. When electrolyte was started to add to the solution, such as 2 mmol of NaCl, the crowded nanorods were developed and with the rise of concentration of NaCl to 4 mmol, the nanorods formation enhanced having larger size and shape as distinguished in Fig. 4(c). When NaCl electrolyte concentration was reached to 6 mmol, very rough, tiny, and dense spherical grains as well as some overgrown clusters were seen. An overgrown cluster was formed due to the coalescence of the particles [36]. Further addition of NaCl concentration was culminated at 8 mmol and exhibited distinctively distributed, clear, and larger sized spherical grains. Thus, the increasing content of NaCl electrolyte changed surface morphologies from nanorods to spherical grains which have potential influences in electrical and optical properties described in the later sections.

The morphology of the $Cu_2O$ nanostructures was sensitive to the concentration of salts added as also reported for CuO [37]. When NaCl concentration was increased gradually, the growth of nanorods also increased but until a limit such as 4 mmol of NaCl addition. These phenomena indicate that NaCl concentration will result in similar morphology of the product and play key roles in controlling the size and shape of the $Cu_2O$ nanostructures. Besides the above-mentioned reasons, the steric hindrance effect caused by salt concentration should also influence the micelle aggregates, and these effects together result in the assemblies of the products. Further studies and work are ongoing to additional research of the mechanisms for the fabrication process caused by the new proposed route.

### 3.3 Electrical Properties:

The electrical resistivities of the samples were measured by using a homemade 4-point collinear probe that was reported in our previous work [23]. Measurements were taken at several regions of the sample under investigation and the results are the representative average values of all measurements summarized in Table-2. Thickness as well as the type of conductivity of films were measured by a similar technique described in ref. [23].

**Table-2: Electrical Properties of the deposited films**

| Sample ID | Thickness (nm) | Sheet resistance (MΩ/square) | Surface resistivity (Ω.cm) | Activation energy $E_a$(eV) | Types of conductivity | Bandgap of samples $E_g$ (eV) | |
|---|---|---|---|---|---|---|---|
| | | | | | | as deposited | annealed |
| S0 | 1350 ± 80 | 33.46 ± 4.23 | 15142 ± 33.84 | 0.16 ± 0.02 | *p*-type | 2.36 | 2.04 |
| S2 | 1270 ± 70 | 2.95 ± 1.14 | 1256 ± 7.98 | 0.21 ± 0.01 | *p*-type | 2.04 | 1.85 |
| S4 | 1610 ± 23 | 1.27 ± 0.11 | 685 ± 0.25 | 0.14 ± 0.02 | *p*-type | 1.96 | 1.96 |
| S6 | 1060 ± 10 | 4.40 ± 1.71 | 1563 ± 1.71 | 0.19 ± 0.02 | *p*-type | 2.0 | 2.02 |
| S8 | 630 ± 10 | 4.82 ± 0.97 | 1018±0.97 | 0.15 ± 0.02 | *p*-type | 2.24 | 1.88 |

From Table-2 the surface resistivity values were found in between 15142 and 685 Ω.cm. Thicknesses were used to calculate the surface resistivity of the samples. The average thickness of the film is 1184 nm. The changes in surface resistivity with thicknesses are shown in Fig. 5(a). It is clearly visible that sample S4 showed the maximum thickness with the minimum resistivity among all the samples. With increasing the content of NaCl electrolyte into the cationic precursor solution, the surface resistivity (S2 – S8) is reduced as compared to the sample deposited without NaCl (S0). The resistivity value dropped to 685 Ω.cm in the sample deposited at 4 mmol NaCl electrolyte (S4) which is about 22 times lower than the sample S0.

The change of surface resistivity with increasing content of NaCl electrolyte may be explained based on the SEM micrographs observed in Fig. 4. As the density, size, and shape of nanorod was increased with increasing the content of NaCl up to 4 mmol as observed from Fig. 4 (a – c), the probability of passing the electrical current through the sample S4 is maximum and consequently the minimum order of surface resistivity (S4, $\rho$ = 685 $\Omega$.cm). Moreover, for the sample deposited at 6 mmol NaCl (S6) the surface was very rough and dense as seen from Fig. 4(d), which may be the cause of a little bit of high resistivity in comparison to sample S4. Furthermore, sample S8 showed lower resistivity than S6 due to the well-distributed and larger spherical grains, which is also understandable by the outlook of the micrographs in Fig. 4(d-e).

The type of conductivity was determined by using the hot-probe method [23]. The temperature dependent surface resistivity ($\rho$) was also measured to determine the activation energy ($E_a$) of the samples. Resistivity value was measured for respective surface temperature of the deposited film in the range between 30 ~ 230 °C. The $E_a$ values were calculated by using the following Arrhenius equation [7]:

$$\rho = \rho_0 \exp\left(\frac{-E_a}{k_B T}\right) \ldots\ldots\ldots (1)$$

$$\text{or, } \log \rho = \left(-\frac{E_a}{1000 k_B}\right)\left(\frac{1000}{T}\right) + \log \rho_0 \ldots\ldots (2)$$

where, $\rho$ is the surface resistivity of the films at a specific temperature, $E_a$ is the activation energy, T is the temperature in Kelvin, $\rho_0$ is the proportionality constant and $k_B$ is the Boltzmann constant (8.617 × $10^{-5}$ eV $K^{-1}$). By plotting $\log \rho$ Vs $\frac{1000}{T}$ as shown in Fig. 5(b), the activation energy $E_a$ is obtained as listed in Table-2. Likewise, the surface resistivity, the activation energy followed the same trend, and the values were found to be 0.14 − 0.21 eV. The measured values were perfectly matched with the reported results [25, 38] and in this study the sample deposited with 4 mmol NaCl electrolyte exhibited the lowest resistivity and activation energy as well as well distributed nanorods. Hence, the sample S4 has preferentially showed good quality than the other deposited $Cu_2O$ samples. And overall, the significant effect of adding NaCl electrolyte in case of depositing $Cu_2O$ thin films have been pronounced.

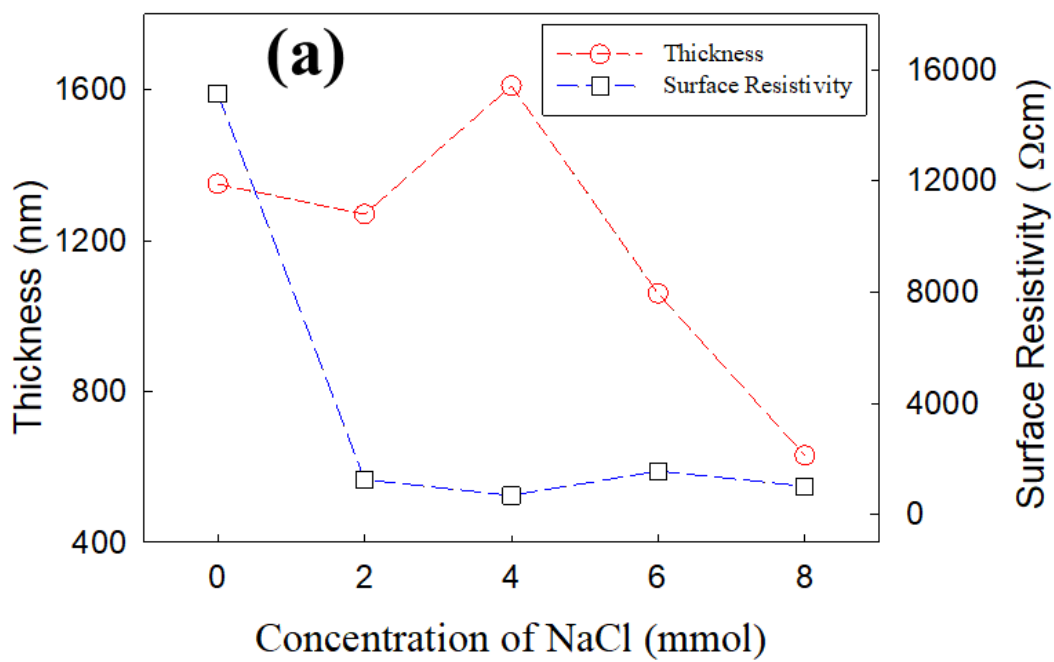

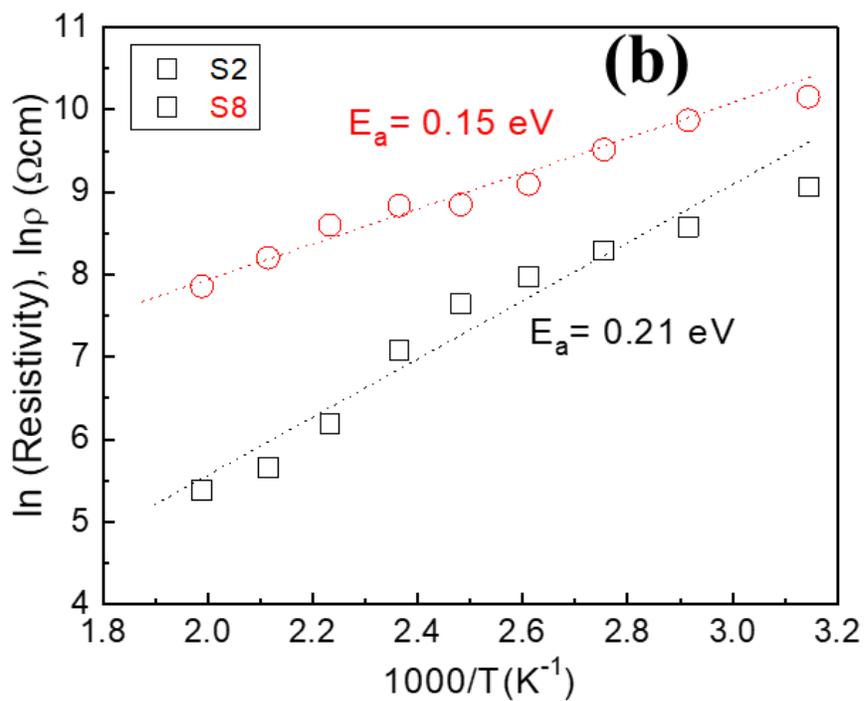

Fig. 5: (a) Variation of surface resistivity with thickness and (b) Temperature dependent activation energy of the samples

## 4.4 Optical Analysis

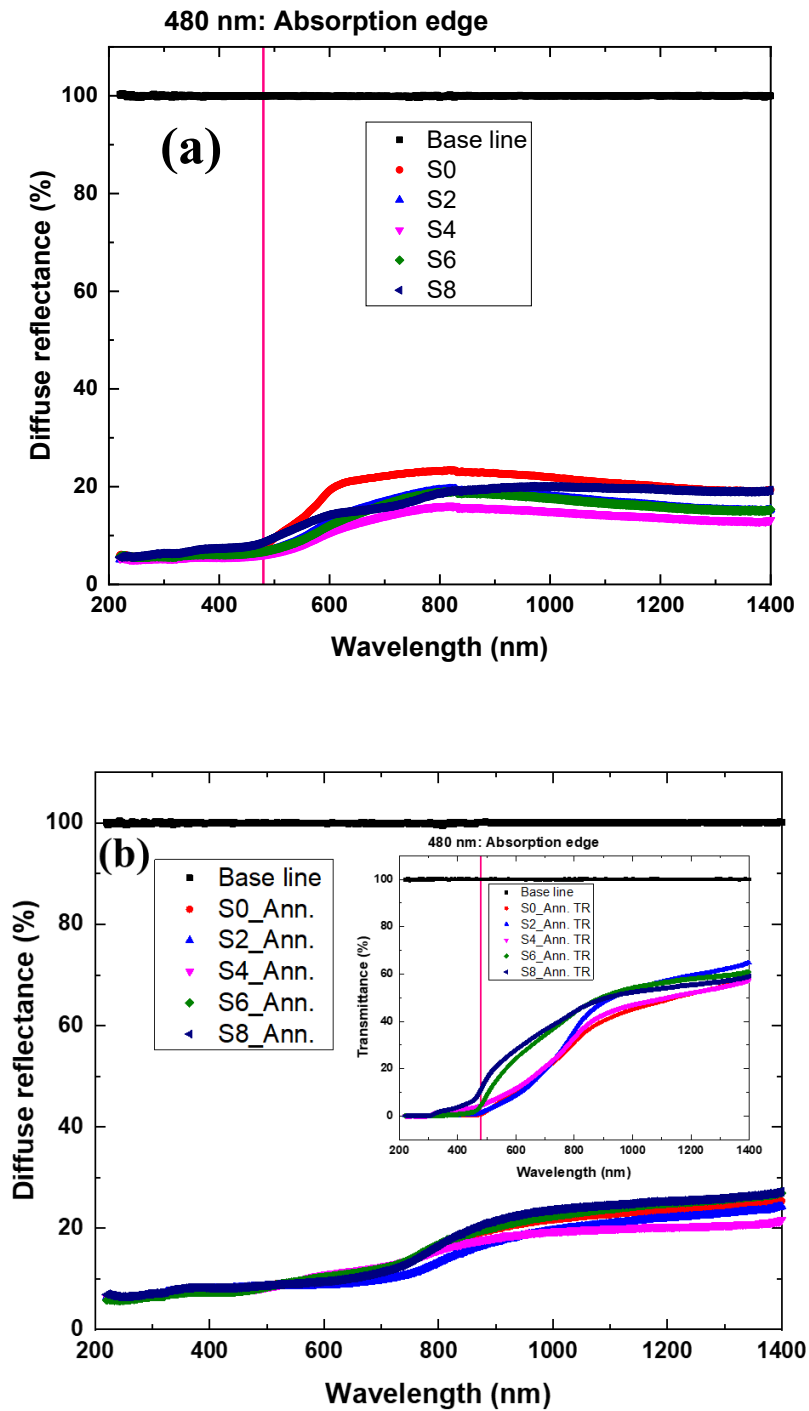

Fig. 6: Diffuse reflectance spectra of (a) as deposited and (b) annealed samples (The absorption edge is drawn at λ≈ 480 nm by pink line and the transmittance data are shown inset of (b) to clarify the phase present in the annealed samples)

To assess the optical characteristics of the samples deposited at various NaCl electrolyte concentrations, the diffuse reflectance spectra of the samples have been taken. UV-Visible-NIR ranges of wavelength 220−1400 nm has been passed through the samples from the respective light sources and the recorded diffuse reflectance spectra are shown in Fig. 6.

From Fig. 6 (a) it is observed that a strong absorption edge occurs at wavelength, λ ≈ 480 nm which corresponds to $Cu_2O$ phase only. No other phase such as CuO is absent which was preliminarily confirmed from both XRD and Raman spectra those were shown in Fig. 2(a−b). In our study, the diffuse reflectance spectra were mainly taken to eliminate any effect that may come from the SLG substrates. The reflectance was seen about 10 ∼ 20%. To investigate the NaCl electrolyte concentration on the optical band gap ($E_g$) of the deposited films Tauc plot was drawn by using reflectance data in correlation with Kubelka- Munk function $F(R_\infty)$ of the following equation [3]:

$$h\nu F(R_\infty)^n = A\,(h\nu - E_g) \dots\dots\dots (3)$$

where, h is the Planck's constant ($6.626 \times 10^{-34}\ Js$), $E_g$ is the optical bandgap, $R_\infty$ is the diffuse reflectance, A is the proportionality constant, $\nu$ is the frequency of the incident photon, n = $\frac{1}{2}$ for indirect band gap semiconductor and n = 2 for direct band gap semiconductor. By plotting $h\nu F(R_\infty)^2$ Vs hν (putting n = 2 for direct band gap $Cu_2O$ semiconductor) and subsequently extrapolated to x-axis; the curves of the following types were found for as deposited and annealed samples, and the respective $E_g$ values are shown in Table -2.

From Table-2, it is seen that for as deposited samples the $E_g$ values were 2.0 ∼ 2.36 eV [39] whereas it was 1.88 − 2.04 eV for annealed samples [24]. In the case of as-deposited samples (Fig. 7(a)), the $E_g$ value dropped from 2.36 to 1.96 eV and then again increased up to 2.24 eV. The sample deposited at 4 mmol NaCl electrolyte showed the lowest bandgap ($E_g$) among all the samples having the smallest surface resistivity (685 Ω.cm) and activation energy (0.14 eV) as well as the well distributed nanorods that observed from Fig. 4(c). On the other hand, air annealing lowered the $E_g$ values those observed from Table-2 and Fig. 7(b) and the value was approximately 1.95 eV. The former reports [24, 40] and current observed $E_g$ values can conclude that air annealing lowered the band gap but did not change $Cu_2O$ phase into CuO up to annealing at 230 °C (Cleared from the inset transmittance absorption spectra shown in Fig. 6 (b)). Hence, air annealing is beneficial to lower the band gap.

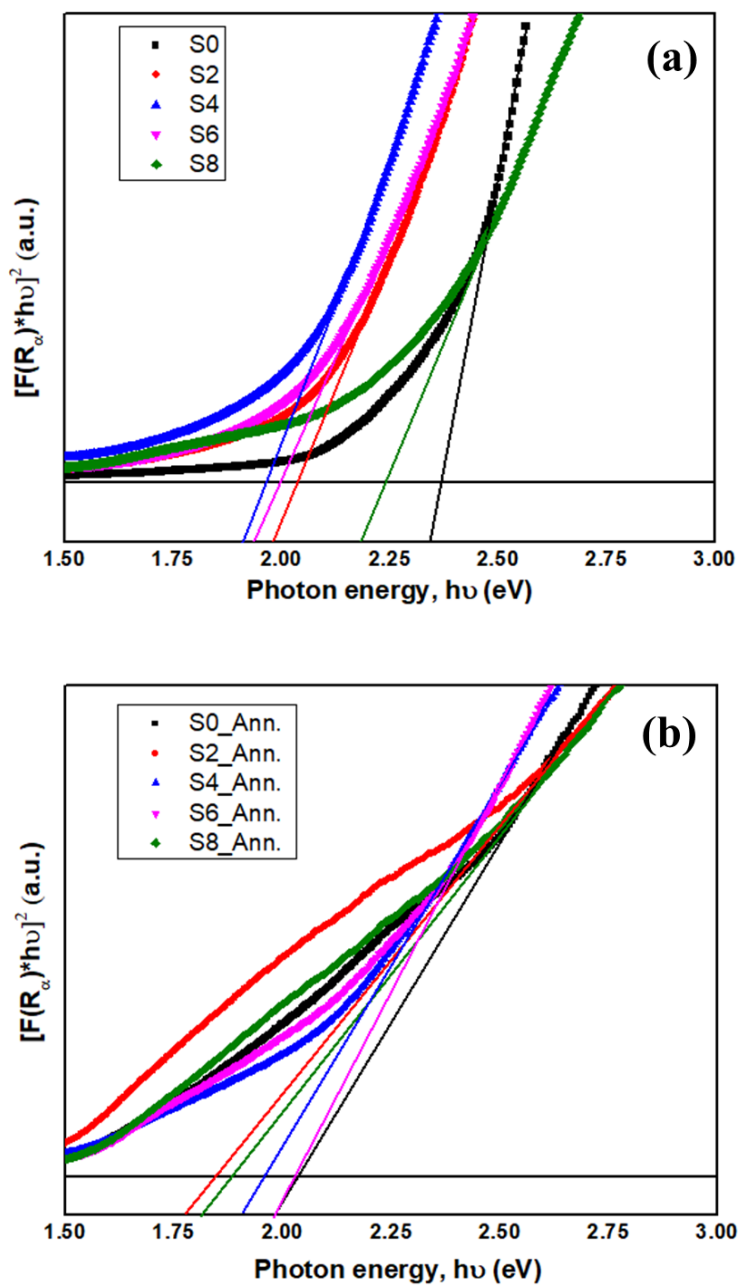

Fig. 7: Optical band gap plot of (a) as deposited and (b) annealed samples deposited at various NaCl electrolyte concentrations

## Conclusion:

In this research, we synthesized $Cu_2O$ nanorod films by the variation of concentration of NaCl on top of simple soda-lime glass by the modified SILAR technique and the structural, electrical, and optical properties of the nanorod films have been investigated. Structural analysis by XRD and Raman validate the polycrystalline pure $Cu_2O$ phase with (111) preferential growth. The SEM micrographs reveal that the deposited films were nanorod structures and closely packed spherical grains, formed with the variation of NaCl concentration. The optical band gap of $Cu_2O$ films estimated by the UV–VIS–NIR spectroscopy was observed to be in the range of 1.88-2.36 eV and consistent with the reported results in the literature. The resistivity of the $Cu_2O$ nanorod films decreased from 15,142 to 685 $\Omega$.cm with the addition of NaCl from 0 to 4 mmol, while the temperature-dependent activation energies of the films were found as about 0.14-0.21 eV. The film optimized in presence of 4 mmol of NaCl demonstrated the lowest resistivity, activation energy and excellent nanorod growth. These results could eventually demand significant attention in the photovoltaic community and research into the development of ecofriendly as well as cost-effective $Cu_2O$ nanorod films-based optoelectronics.

## Data accessibility:

Our data are deposited at Dryad as (DOI): https://doi.org/10.5061/dryad.37pvmcvmm

## Acknowledgements:

M.A.M.P, M.A.R and M.A.H happily acknowledge the lab facilities at Physical Chemistry Research Laboratory and funding of Comilla University, Cumilla and UGC, Bangladesh. M.A.H, S.F.U.F. and N.I.T. gratefully acknowledge the experimental support of the Energy Conversion and Storage Research (ECSR) Section, Industrial Physics Division (IPD), BCSIR Laboratories, Dhaka under the scope of R&D project # 100-FY2017-2021. S.F.U.F. also acknowledges the support of RSC research grant # R20-3167 for ECSR,

IPD. All the authors thankfully acknowledge the experimental support from the Optoelectronics laboratory of Saga University, Saga, Japan.